\documentclass{webofc}
\usepackage[varg]{txfonts}   
\def\be{\begin{equation}}
\def\ee{\end{equation}}
\def\bea{\begin{eqnarray}}
\def\eea{\end{eqnarray}}
\def\dd#1{\frac{\mathrm{d}^2#1}{(2\pi)^2}}
\def\ddd#1{{\mathrm{d}^2#1}}

\begin{document}
\title{The Weizs\"acker-Williams distribution of linearly polarized
  gluons (and its fluctuations) at small x}
%
%

\author{\firstname{Adrian} \lastname{Dumitru}\inst{1,2,3}
  \fnsep\thanks{\email{adrian.dumitru@baruch.cuny.edu}} \and
  \firstname{Vladimir} \lastname{Skokov}\inst{4}
  \fnsep\thanks{\email{vskokov@quark.phy.bnl.gov}}
}

\institute{Department of Natural Sciences, Baruch College, CUNY,
17 Lexington Avenue, New York, NY 10010, USA
\and
The Graduate School and University Center, The City
University of New York, 365 Fifth Avenue, New York, NY 10016, USA
\and
Physics Department, Brookhaven National Lab, Upton, NY 11973, USA
\and
RIKEN/BNL Research Center, Brookhaven National
  Laboratory, Upton, NY 11973, USA
          }

\abstract{The conventional and linearly polarized
  Weizs\"acker-Williams gluon distributions at small $x$ are defined
  from the two-point function of the gluon field in light-cone
  gauge. They appear in the cross section for dijet production in deep
  inelastic scattering at high energy. We determine these functions in
  the small-$x$ limit from solutions of the JIMWLK evolution equations
  and show that they exhibit approximate geometric scaling. Also,
  we discuss the functional distributions of these WW gluon
  distributions over the JIMWLK ensemble at rapidity $Y\sim
  1/\alpha_s$. These are determined by a 2d Liouville action for the
  logarithm of the covariant gauge function $g^2 \mathrm{tr}\,A^+(q)A^+(-q)$.
  For transverse momenta on the order of the saturation scale we
  observe large variations across configurations (evolution
  trajectories) of the linearly polarized distribution up to several
  times its average, and even to negative values.}
\maketitle
\section{Introduction}
\label{sec:intro}

Dijet production in deep-inelastic $\gamma^*-A$ scattering at high
energy can provide insight into the gluon fields of the nucleus in the
regime of strong, non-linear fields~\cite{Aschenauer:2017jsk}. At
leading order a $q\bar{q}$ dijet is produced. Denote the average
transverse momentum of the jets as $\vec{P}=(\vec{k}_1-\vec{k}_2)/2$
and the transverse momentum imbalance as $q=\vec{k}_1+\vec{k}_2$,
where $\vec{k}_1$ and $\vec{k}_2$ are the transverse momenta of the
two jets.  In the ``correlation limit'' of roughly back to back
jets~\cite{Dominguez:2011wm} one has $P^2\gg q^2$. In this limit the
leading contribution (in powers of $q^2/P^2$) to the cross section can
be obtained from Transverse Momentum Dependent (TMD) factorization. It
predicts a distribution $xh^{(1)}(x,q^2)$ for linearly polarized
gluons in an unpolarized target~\cite{Mulders:2000sh,Meissner:2007rx}
which gives rise to $\sim\cos 2\phi$ azimuthal anisotropies in dijet
production~\cite{Boer:2009nc,Metz:2011wb,DQXY}, as well as in other
processes~\cite{Boer:2010zf,Qiu:2011ai,Lansberg:2017tlc}. $\phi$ is
the angle between the transverse momentum vectors $\vec{P}$ and
$\vec{q}$ (in a frame where neither the $\gamma^*$ nor the hadronic
target carry transverse momentum). The isotropic contribution to the
dijet cross section is proportional to the conventional
Weizs\"acker-Williams (WW) gluon distribution $xG^{(1)}(x,q^2)$:
\begin{eqnarray}
E_1E_2
\frac{d\sigma ^{\gamma _{T}^{\ast }A\rightarrow
    q\bar{q}X}}{d^3k_1d^3k_2 d^2b}
&=&\alpha _{em}e_{q}^{2}\alpha _{s}\delta \left( x_{\gamma ^{\ast
}}-1\right) z(1-z)\left( z^{2}+(1-z)^{2}\right) \frac{\epsilon _{f}^{4}+%
{P}^{4}}{({P}^{2}+\epsilon _{f}^{2})^{4}}
\notag \\ 
&&
\quad \quad \quad \quad \quad \quad 
\times \left[ xG^{(1)}(x,q)-\frac{2\epsilon _{f}^{2}{P}%
^{2}}{\epsilon _{f}^{4}+{P}^{4}}\cos
  \left(2\phi\right)xh^{(1)}(x,q)\right] ~,
\label{dipoledis2} \\
E_1E_2
\frac{d\sigma ^{\gamma _{L}^{\ast }A\rightarrow q\bar{q}X}}{d^3k_1d^3k_2 d^2b}
&=&\alpha _{em}e_{q}^{2}\alpha _{s}\delta \left( x_{\gamma ^{\ast
}}-1\right) z^{2}(1-z)^{2}\frac{8\epsilon _{f}^{2}{P}^{2}}{(
{P}^{2}+\epsilon _{f}^{2})^{4}}  \notag \\
&&
\quad \quad \quad \quad \quad \quad 
\times \left[ xG^{(1)}(x,q)+\cos \left(2
  \phi\right)xh^{(1)}(x,q)\right]~. 
\label{eq:dipoledisl2}
\end{eqnarray}
$z$ and $1-z$ are the momentum fractions of the quark and anti-quark,
respectively, and $\epsilon_f^2=z(1-z)Q^2$ (for massless quarks) with
$Q^2$ the virtuality of the photon.  Clearly, positivity of the cross
section imposes the upper bound $|xh^{(1)}(x,q^2)| \le
xG^{(1)}(x,q^2)$. Note that even though $P$ is taken to be the hard
scale in the process, which can be greater than the saturation scale
of the nucleus, that nevertheless the WW gluon distributions are
probed at the much smaller momentum imbalance scale $q$. Therefore,
the process can indeed provide information on these gluon
distributions in the dense regime at $q<Q_s$. The WW gluon
distributions also determine the divergence of the Chern-Simons
current at the initial time in relativistic heavy-ion
collisions~\cite{Lappi:2017skr} even though they are not the gluon
distributions which enter the cross section for gluon production~\cite{Kharzeev:2003wz}.

In the Color Glass Condensate (CGC) framework at small~$x$ the gluon
fields are described by Wilson lines. They are path ordered
exponentials in the strong color field of the target, and cross
sections for different observables can be related to correlators of
the Wilson lines.  The Wilson line is a path ordered exponential of
the covariant gauge field, whose largest component is $A^+$:
\be \label{eq:V_rho} U(\vec{x}) = \mathbb{P} \exp\left\{ ig \int d x^-
A^+(x^-,\vec{x}) \right\}.  
\ee 
The WW unintegrated gluon
distribution~\cite{Dominguez:2011wm,Kharzeev:2003wz,Dominguez:2011br},
on the other hand, is defined in terms of the light cone gauge
($A^+=0$) field; it can be obtained by a gauge transformation
\be \label{eq:E_WW}
    {A}^i(\vec{x}) =
\frac{1}{ig} U^\dagger(\vec{x}) \, \partial_i U(\vec{x}) ~.  
\ee
The trace (or the traceless part) of the two-point correlator of the
light cone gauge field
\begin{equation}
x G^{ij}_{\rm WW} (x, \vec{q}) =  \frac{1}{A_\perp} \frac{1}{2\pi}
\left< A^i_a(\vec{q}) A^j_a(-\vec{q}) \right>
\label{eq:xGij}
\end{equation}
defines $xG^{(1)}(x,q^2)$ and $xh^{(1)}(x,q^2)$ introduced above:
\begin{equation}
x G^{ij}_{\rm WW}(x,q^2)  = \frac{1}{2} \delta^{ij} xG^{(1)}(x,q^2)
+\frac{1}{2} \left(2 \frac{q^i q^j}{q^2} - \delta^{ij}\right)
xh^{(1)}(x,q^2) ~.   \label{eq:xGij_xGxh} 
\end{equation}
$A_\perp$ in eq.~(\ref{eq:xGij}) denotes a transverse area over which
the gluon distributions have been averaged over.

\section{The WW gluon distributions at small $x$}
\label{sec-WW}

\begin{figure*}
\centering
\includegraphics[width=0.45\linewidth]{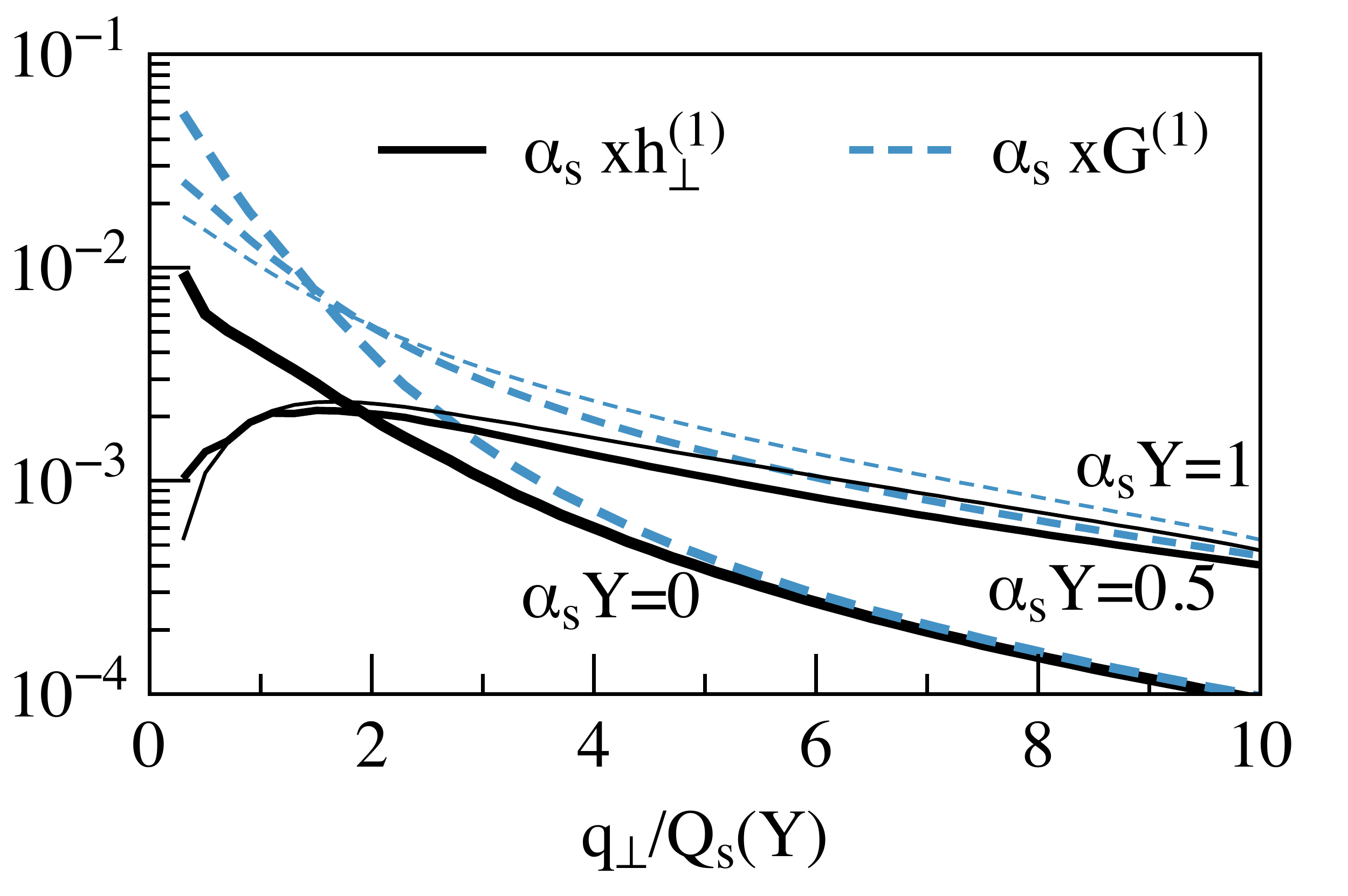}
\includegraphics[width=0.45\linewidth]{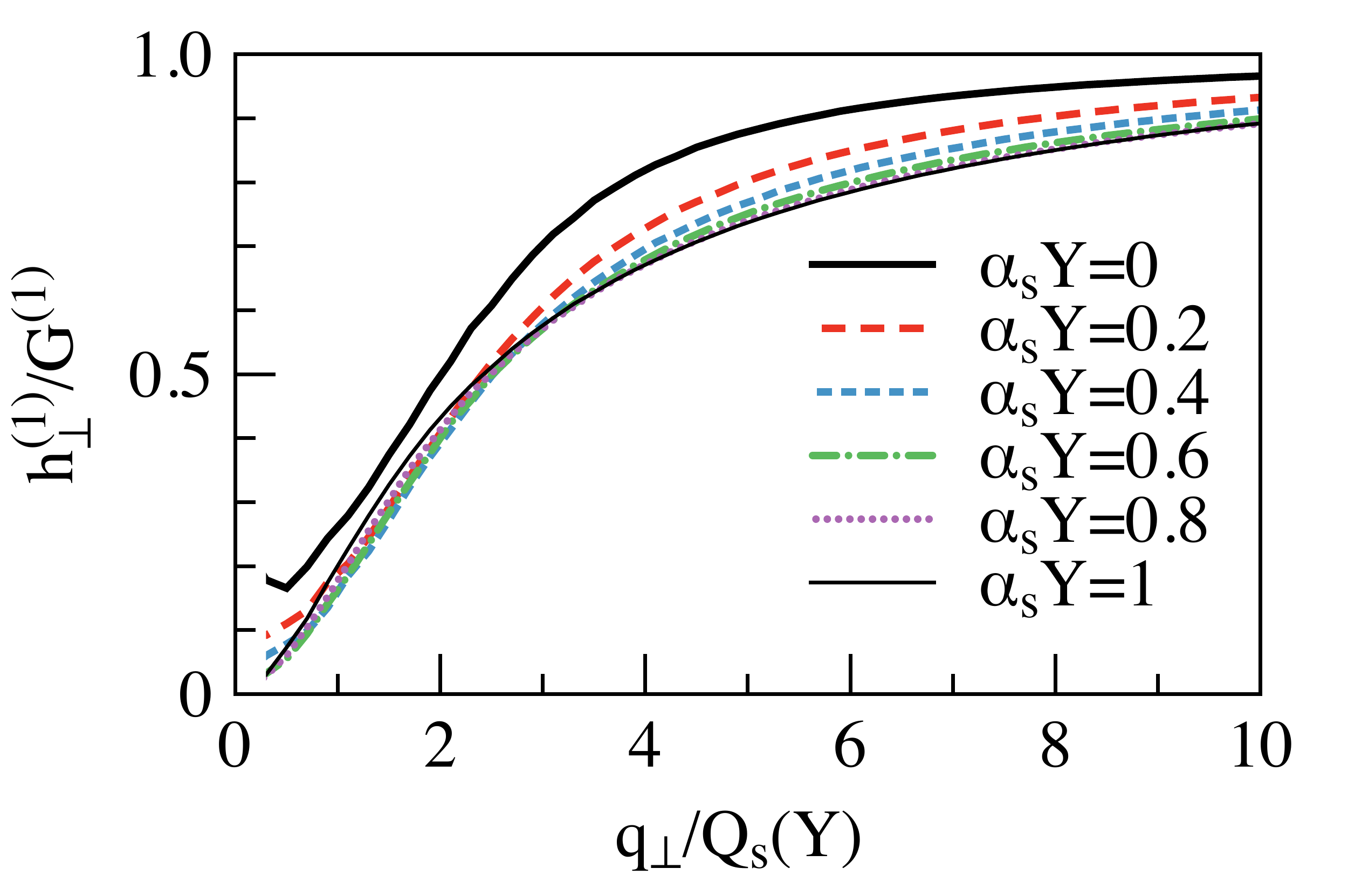}
\caption{$xG^{(1)}(x,q^2)$ and $xh^{(1)}(x,q^2)$ WW gluon
  distributions versus transverse momentum $q$ at different rapidities
  $Y$. $Q_s(Y)$ is the saturation momentum. The curves correspond to
  evolution at fixed $\alpha_s$.}
\label{fig:xGxh}       
\end{figure*}
These WW gluon distributions have been obtained at small $x$ (high
rapidity\footnote{$Y=\log x_0/x$ where $x_0$ determines the onset of
  small-$x$ evolution; it is typically taken to be $x_0=10^{-2}$.}) by
a numerical solution of the JIMWLK evolution equations~\cite{jimwlk}
in ref.~\cite{Dumitru:2015gaa}, shown in fig.~\ref{fig:xGxh}. At high
transverse momentum one finds that $xh^{(1)}(x,q^2)\to
xG^{(1)}(x,q^2)$. This is easy to understand from the fact that in the
dilute limit the classical light-cone gauge field is given by $A^i(q)=
ig(q^i/q^2)\rho(q)$ where $g\rho(q)$ denotes the color charge density
of the sources. For $A^i(q)\sim q^i$
eqs.~(\ref{eq:xGij},\ref{eq:xGij_xGxh}) give $xh^{(1)}(x,q^2)=
xG^{(1)}(x,q^2)$. Thus, the saturation of the above-mentioned bound on
the distribution of linearly polarized gluons at high transverse
momentum is a generic consequence of the dilute semi-classical field
limit.  On the other hand, at low $q$ one has $xh^{(1)}(x,q^2)/
xG^{(1)}(x,q^2)\approx0$ implying that the angular dependence of the
cross section~(\ref{eq:dipoledisl2}) is weaker.  For more detailed
predictions of $\langle\cos 2\phi\rangle$ obtained with the small-$x$
WW gluon distributions see ref.~\cite{Dumitru:2015gaa}.

The numerical solutions also indicate that the WW gluon distributions
approach scaling functions $xh^{(1)}(x,q^2)=xh^{(1)}(q^2/Q_s^2(x))$,
$xG^{(1)}(x,q^2) =xG^{(1)}(q^2/Q_s^2(x))$, at high rapidity. This is
known as geometric scaling and has been discussed originally in the
context of the dipole forward scattering amplitude (resp.\ the
$\gamma^* -p$ total cross section)~\cite{Stasto:2000er}.  Geometric
scaling of the WW distributions can be motivated from a Gaussian
approximation to JIMWLK~\cite{Iancu:2002aq}. In this approximation,
and also taking $N_c\gg1$ for simplicity, they
can be written in terms of the two-point function
$\Gamma(r)$ of $A^+$ as~\cite{Dumitru:2016jku}
\bea xh^{(1)}(x,q^2)
&=& \frac{4 N_c}{\alpha_s (2\pi)^3} \int dr r^3
J_2(qr) \left(1-[S(r^2)]^2\right) \frac{\Gamma''(r^2)}{\Gamma (r^2)} \\
xG^{(1)}(x,q^2) &=& \frac{4 N_c}{\alpha_s (2\pi)^3}
\int dr r J_0(qr)
\left(1-[S(r^2)]^2\right) \left( \frac{\Gamma'(r^2)}{\Gamma (r^2)}
+ r^2 \frac{\Gamma''(r^2)}{\Gamma (r^2)}
\right)~.
\eea
$S(r^2)=\exp\left(-(1/2) C_F \Gamma(r^2)\right)$ is the S-matrix for a
dipole of size $r$. At the JIMWLK fixed point $\Gamma(r^2)$, $r^2
\Gamma'(r^2)$, and $r^4 \Gamma''(r^2)$ are in fact
functions of $r^2 Q_s^2(x)$ only, rather than functions
of both $r$ and $x$. From the above expressions for $xh^{(1)}(x,q^2)$
and $xG^{(1)}(x,q^2)$ it follows that these functions then satisfy
geometric scaling (also see ref.~\cite{Marquet:2016cgx}).

\section{The JIMWLK weight functional and the constraint effective
  potential for $g^2\mathrm{tr}\, |A^+(q)|^2$}
\label{sec:JIMWLK_W[A+]}

Expectation values of observables at small $x$ are computed by i)
expressing the observable as a functional $O[A^+]$ of the covariant
gauge field, ii) and averaging over the random semi-classical fields
$A^+$ with the weight $W_Y[A^+]$:
\be
\langle O\rangle = \frac{1}{Z} \int {\cal D}A^+(q) \, W_Y[A^+(q)] \,
O[A^+(q)]~~~~,
~~~~ Z = \int {\cal D}A^+(q) \, W_Y[A^+(q)] ~. \label{eq:<O>}
\ee
Note that $A^+(q)$ is the soft semi-classical field which solves the
Poisson equation, $A^+(q) = g\rho(q)/q^2$, with $g\rho(q)$ the random,
effective color charge density that is the source of the soft gluon
field. Hence, the above average over $A^+$ can also be written as an
average over $\rho$.

The weight $W_Y[A^+(q)]$ for a given configuration $A^+(q)$ is
determined by the solution of the JIMWLK functional RG
equation~\cite{jimwlk}. The exact solution can only be obtained
numerically. However, a non-local (in coordinate space) Gaussian
mean-field approximation for $W_Y[\rho]$ has been proposed, see first
reference in~\cite{Iancu:2002aq}, which reproduces the proper gluon
distribution $g^2\mathrm{tr}\, A^+(q)\,A^+(-q)$ both at small ($q^2\ll
Q_s^2$) as well as at high ($q^2\gg Q_s^2$) transverse momentum:
\be
W_G[A^+] = e^{-S_G[A^+]}~~~,~~~
S_G[A^+] = \int \dd q \, q^4 \frac{\mathrm{tr}\, A^+(q)\, A^+(-q)}
{g^2\mu^2(q^2)}~.
\ee
For simplicity we restrict here to high transverse momentum where
\be \label{eq:mu2_q2}
\mu^2(q^2) \simeq \mu_0^2 \left(\frac{q^2}{Q_s^2}\right)^{1-\gamma}~,
\ee
with $\gamma\simeq0.64$ an anomalous
dimension~\cite{Mueller:2002zm}. $Q_s^2$ and $\mu_0^2\sim A^{1/3}$ are
evaluated at the rapidity of interest but we will not spell out this
dependence on $Y$ explicitly.

The expectation value of $O[A^+]$ written in eq.~(\ref{eq:<O>}) is an
average over all configurations of $A^+(q)$. However, one may be
interested in evaluating $\langle O\rangle$ over a subclass of
configurations, for example those with a high (or low) number of
gluons, or with an unusual transverse momentum distribution of
gluons. To that end we introduce the constraint effective potential
for $X(q)=g^2 \mathrm{tr}\, |A^+(q)|^2$ by integrating over
configurations at fixed $X(q)$~\cite{Dumitru:2017cwt}:
\bea
Z &=& \int {\cal D}X(q) \, e^{-V_\mathrm{eff}[X(q)]}~,\label{eq:Z(X)}\\
e^{-V_\mathrm{eff}[X(q)]} &=& \int {\cal D}A^+(q) \, W_Y[A^+(q)] \,
\delta(X(q)-g^2 \mathrm{tr}\, |A^+(q)|^2)~.
\eea
For a Gaussian theory the integral over configurations at fixed $X(q)$
is easy to compute, and the resulting effective potential is~\cite{Dumitru:2017cwt}
\be \label{eq:Veff_A+A+}
V_\mathrm{eff}[X(q)] = \int \dd q \left[\frac{q^4}{g^4\mu^2(q)}X(q) -
  \frac{1}{2} A_\perp N_c^2\log X(q)\right]~,
\ee
$A_\perp$ is the transverse area over which
\be
\mathrm{tr}\, |A^+(q)|^2 = \int_{A_\perp} \ddd b \int \ddd r \, e^{-i\vec q
  \cdot \vec r} A^+(\vec b-\vec r/2) A^+(\vec b+\vec r/2)
\ee
has been integrated.
The stationary point of $V_\mathrm{eff}[X]$ determines the extremal
gluon distribution
\be
X_s(q)=\frac{1}{2} N_c^2 A_\perp \frac{g^4\mu^2(q)}{q^4}~.
\ee
$X_s(q)$ is the most likely gluon distribution function rather than
the average. However, in the large-$N_c$ limit it is
equal to the expectation value of $\langle g^2 \mathrm{tr}\,
|A^+(q)|^2\rangle$. Away from the extremal solution, the potential
$V_\mathrm{eff}[X]$ provides insight into the distribution of
functions $X(q)$ about the extremum. This distribution is determined
by a ``linear minus logarithmic'' rather than by a polynomial potential.

It will be convenient for what follows to describe deviations from
$X_s(q)$ by multiplying with $\eta(q)$ rather than by adding $\delta
X(q)$. Hence, we introduce the function $\eta(q)$ through $X(q) = X_s(q)\,
\eta(q)$.  A fluctuation from the extremal field $X_s(q)$ has action
\bea
V_\mathrm{eff}[\eta(q)]
= \frac{1}{2} N_c^2 A_\perp \int \dd q \left[ \eta(q) -1 -
  \log\eta(q)\right]~.
\label{eq:Veff_eta}
\eea
Note that $X(q)=g^2\mathrm{tr}\, |A^+(q)|^2$ is a positive definite
function and so is $\eta(q)$. We can therefore perform another field
redefinition to introduce $\phi(q)$ through $e^{\phi(q)} = \eta(q)$ so
that
\bea
V_\mathrm{eff}[\phi(q)] =
\frac{1}{2} N_c^2 A_\perp \int \dd q \left[ e^{\phi(q)} - \phi(q) -1\right]~.
\label{eq:Veff_Liouville}
\eea
Thus, we found that it is a Liouville action in two dimensions which
describes the distribution of $\log \mathrm{tr}\, |gA^+(q)|^2$
(relative to the average gluon distribution) in a Gaussian
approximation to JIMWLK.

The action for the most likely distribution function $X_s(q)$ is of
order $N_c^2 A^{1/3}$ (times zero, in dimensional regularization). So
is the action for $X(q)=X_s(q) \eta(q)$ if $\eta(q) = {\cal
  O}(1)$. Our discussion is restricted to the distribution of
functions $X(q)$ which exhibit longitudinal coherence and are of order
$N_c^2 A^{1/3}$. The small-$x$ power counting assumes $g^4
A^{1/3}={\cal O}(1)$~\cite{Kovchegov:1999ua}, and so $X(q)\sim
(A^{1/3})^0$ would correspond to a higher order correction in the
coupling.

Knowing $V_\mathrm{eff}$ we can now evaluate the suppression probability
for a modification of the gluon distribution such as
\be \label{eq:eta(q)}
\eta(q) = 1 + \eta_0 \left(\frac{g^4\mu_0^2}{q^2}\right)^a\,
\Theta\left(q^2-\Lambda^2\right) \Theta\left(Q^2-q^2\right)~.
\ee
$\eta_0$ determines the amplitude of the distortion, $Q^2$ and
$\Lambda^2>Q_s^2$ determine its support, and the parameter $a$ specifies the
spectral shape.
The action for such $\eta(q)$ when $Q^2\gg\Lambda^2$ is
\be \label{eq:Delta_S_eta}
V_\mathrm{eff}[\eta(q)] \simeq \frac{1}{8\pi} N_c^2 A_\perp \, g^4\mu_0^2\, \eta_0
\times
\begin{cases}
  \frac{1}{1-a} \left(\frac{Q^2}{g^4\mu_0^2}\right)^{1-a} & (a<1)~, \\
  \log\frac{Q^2}{\Lambda^2} & (a=1)~,\\
 \frac{1}{a-1}\left(\frac{g^4\mu_0^2}{\Lambda^2}\right)^{a-1} & (a>1)~.
\end{cases}
\ee
Hence, we find that a harder than average gluon distribution ($a<0$)
over $\Lambda<q<Q$ comes at a high price since
$V_\mathrm{eff}[\eta(q)] \sim (Q^2/g^4\mu_0^2)^{1-a}$. On the other
hand, gluon distributions which drop substantially faster than the
most likely one (i.e.\ $X_s(q)$) correspond to $a\ge 1$, and such
fluctuations can extend to high transverse momentum. For a more
detailed discussion of the shape of the gluon distribution in the
presence of a high (or low) gluon multiplicity ``trigger'' we refer to
ref.~\cite{Dumitru:2017cwt}.

\section{The functional distribution of WW gluon distributions over the
  JIMWLK ensemble}
\label{sec:JIMWLK-xGxh}

In sec.~\ref{sec-WW} we discussed the WW gluon distributions averaged
over the entire JIMWLK ensemble $W_Y[A+]$. In this section we discuss
the distributions of these functions over the JIMWLK ensemble.

To obtain some basic analytic insight we write the expansion of $g^2
\mathrm{tr}\, A^i(q) A^j(-q)$ to fourth order in $A^+(q)$ obtained via
eq.~(\ref{eq:E_WW}):
\bea
\delta^{ij} \mathrm{tr}\, A^i(q) A^j(-q) &=&
   \frac{1}{2} q^2 A^{+a}(q) A^{+a}(-q)  \nonumber\\
   && \hspace{-4.5cm}
   + \frac{g^2}{8} f^{abe} f^{cde} \left(\frac{q^n q^m}{q^2}-\delta^{nm}\right)
     \int \dd k k^n A^{+a}(q-k) A^{+b}(k) 
     \int \dd p p^m A^{+c}(-q-p) A^{+d}(p) ~,\\
\left(2\frac{q^i q^j}{q^2}-\delta^{ij}\right) \mathrm{tr}\, A^i(q) A^j(-q) &=&
   \frac{1}{2} q^2 A^{+a}(q) A^{+a}(-q)  \nonumber\\
   && \hspace{-4.5cm}
   - \frac{g^2}{8} f^{abe} f^{cde} \left(\frac{q^n q^m}{q^2}-\delta^{nm}\right)
     \int \dd k k^n A^{+a}(q-k) A^{+b}(k) 
     \int \dd p p^m A^{+c}(-q-p) A^{+d}(p) ~.     \label{eq:xh_config}
 \eea
In the weak-field limit the first term in these expansions dominates
and the two WW gluon distributions are equal, configuration by
configuration. The correction at fourth power in $A^+$ generates a
``splitting''. We can perform an average over a Gaussian ensemble by
summing the two non-vanishing Wick contractions using
\be \label{eq:<A+A+>}
\left<A^{+a}(q) \, A^{+b}(k)\right> = \delta^{ab} \, (2\pi)^2 \delta(q+k)
\frac{g^2\mu^2(q^2)}{q^4}~.
\ee
At order $N_c^2$ this leads to
\be  \label{eq:deltaij_<AiAj>}
\delta^{ij} \, g^2 \left<\mathrm{tr}\, A^i(q) A^j(-q) \right> =
\frac{1}{2} N_c^2 A_\perp \frac{g^4\mu^2(q)}{q^2} + N_c^3 A_\perp
\frac{g^8}{4} \int \dd k k^2 \left[1-(\hat{k}\cdot\hat{q})^2\right]
\frac{\mu^2(q-k)}{(q-k)^4} \frac{\mu^2(k)}{k^4}~.
\ee
The result for the average of eq.~(\ref{eq:xh_config}) is the same
except that the sign of the second term is negative. Thus, one may
wonder if the linearly polarized distribution could take negative
values\footnote{This function does not have a ``gluon density'' /
  probability interpretation and so it needs not be positive
  definite.}. It is clear that at high $q^2$ the correction is $\sim
1/q^2$ power suppressed as compared to the leading contribution. This
suppression ensures that $xh^{(1)}(x,q^2)$, averaged over all
configurations, is a positive definite function (as seen in
fig.~\ref{fig:xGxh}). 

Instead of averaging over all JIMWLK configurations we can use the
approach from the previous section to integrate over all
configurations at fixed $X(q)=g^2 \mathrm{tr}\, |A^+(q)|^2$. To do so,
instead of using eq.~(\ref{eq:<A+A+>}) we make the final integration
over $X(q)$ explicit:
\be \label{eq:<A+A+>_X}
\left<A^{+a}(q) \, A^{+b}(k)\right> = \delta^{ab} \, (2\pi)^2 \delta(q+k)
\frac{g^2\mu^2(q^2)}{q^4}
\int {\cal D}X(\ell) \, e^{-V_\mathrm{eff}[X(\ell)]}\, \frac{X(q)}{X_s(q)}~.
\ee
We can then rewrite eq.~(\ref{eq:deltaij_<AiAj>}) as follows:
\bea
\delta^{ij} \, g^2 \left<\mathrm{tr}\, A^i(q) A^j(-q) \right> &=&
q^2 \int {\cal D}X(\ell) \, e^{-V_\mathrm{eff}[X(\ell)]}\,
X(q) \nonumber\\
& & \hspace*{-1cm}+ \frac{1}{N_cA_\perp} 
\int \dd k k^2 \left[1-(\hat{k}\cdot\hat{q})^2\right]
\int {\cal D} X(\ell) \, e^{-V_\mathrm{eff}[X(\ell)]}\, 
X(q-k)X(k)~.
\label{eq:deltaij_<AiAj>_X}
\eea
As before, replacing the projector $\delta^{ij}$ by 
$2{q^i q^j}/{q^2}-\delta^{ij}$ reverses the sign of the last term. It
is evident that for some functions $X(q)$ which contribute to
the integral the correction in this last expression may be greater
than the ``leading'' contribution. These configurations overcome the
power suppression discussed above which arises at the saddle point of
the integral; also, they give linearly polarized gluon distributions
which are negative in some range of transverse momentum.

We now show some numerical results obtained by Monte-Carlo sampling of
the JIMWLK functional $W_Y[A^+]$~\cite{Dumitru:2017cwt} for $N_c=3$
colors and fixed $\alpha_s$. We evaluate the WW gluon distributions on
each configuration. They have to be integrated over a finite patch in
impact parameter space greater than the inverse transverse
momentum. We take
\be
P^{ij} \int \ddd x \ddd y \, e^{-i q\cdot (x-y)} e^{- (x^2+y^2)/2R^2} \,
g^2 \mathrm{tr}\, A^i(x) A^j(y)~,
\ee
with $R=2/Q_s(Y)$ and $q>Q_s(Y)$ the transverse momentum
scale. $P^{ij}$ denotes one of the two projectors mentioned
above. This expression factorizes into a product of two Fast Fourier Transforms
which can be evaluated numerically very efficiently.

\begin{figure}[h]
\centering
\includegraphics[width=0.9\linewidth,clip]{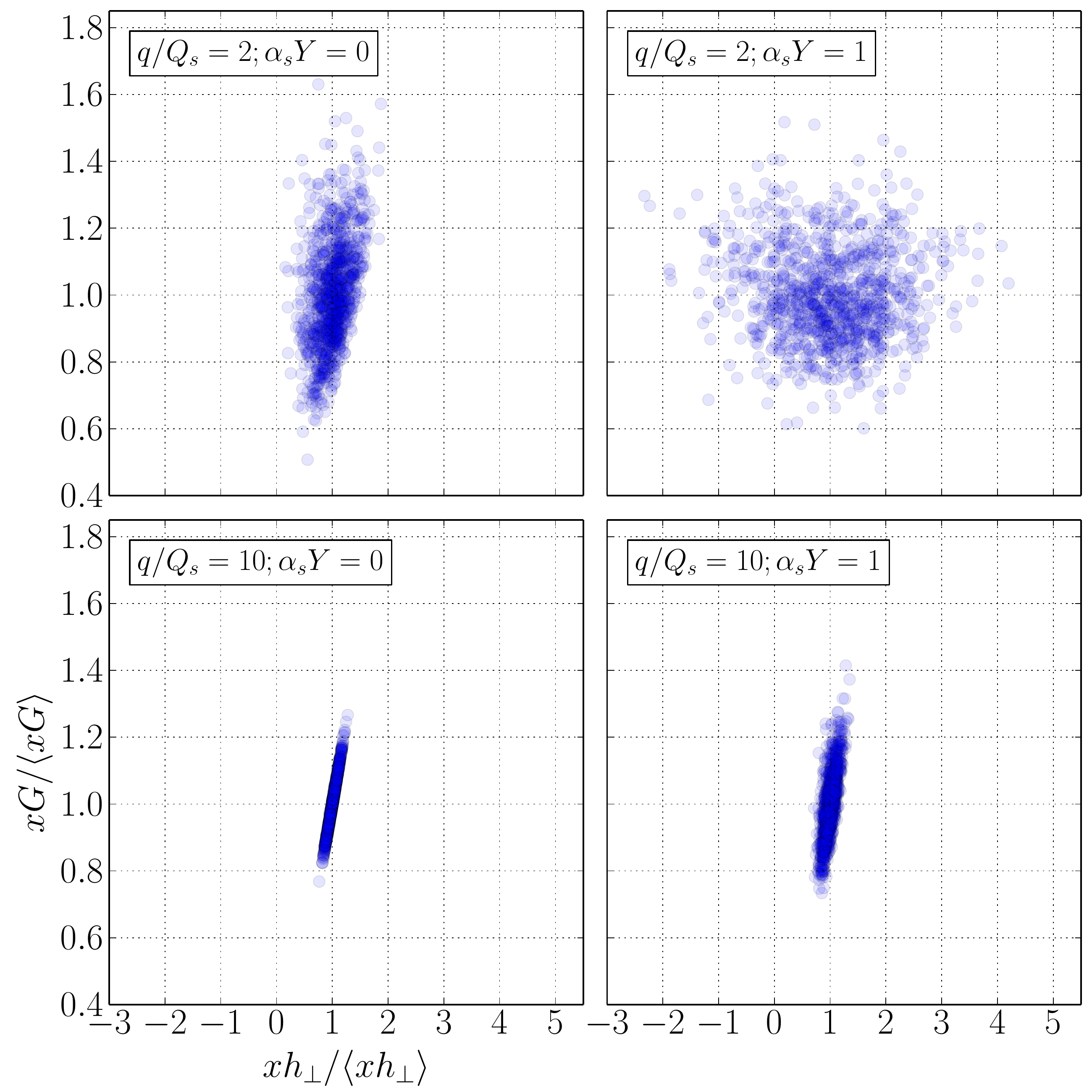}
\caption{$xG^{(1)}(x,q^2)$ and $xh^{(1)}(x,q^2)$ WW gluon
  distributions for 1000 individual field configurations, evaluated at
  $q=2Q_s(Y)$ or $q=10Q_s(Y)$, respectively.  Notice the different
  scales on the horizontal and vertical axes.  Left: MV model initial
  condition; Right: JIMWLK ensemble at $\alpha_s Y=1$.}
\label{fig:xGxh_ensemble}       
\end{figure}
In fig.~\ref{fig:xGxh_ensemble} we show the $xG^{(1)}(x,q^2)$ and
$xh^{(1)}(x,q^2)$ WW gluon distributions for individual
configurations, relative to their average. For high transverse
momentum far above $Q_s(Y)$ we observe, as expected, that the two
functions are essentially equal, even for individual
configurations. For $q=2Q_s(Y)$ on the other hand the relative
fluctuations of the linearly polarized distribution are much greater
than those of the conventional WW distribution. For some
configurations $xh^{(1)}(x,q^2)$ can take values up to several times
its average while other evolution trajectories lead to negative
values. This is an effect of evolution to small $x$ since
$xh^{(1)}(x,q^2)<0$ at $q=2Q_s(Y)$ does not occur at $Y=0$ once in
$10^4$ configurations.

\begin{figure}[h]
\centering
\includegraphics[width=0.9\linewidth,clip]{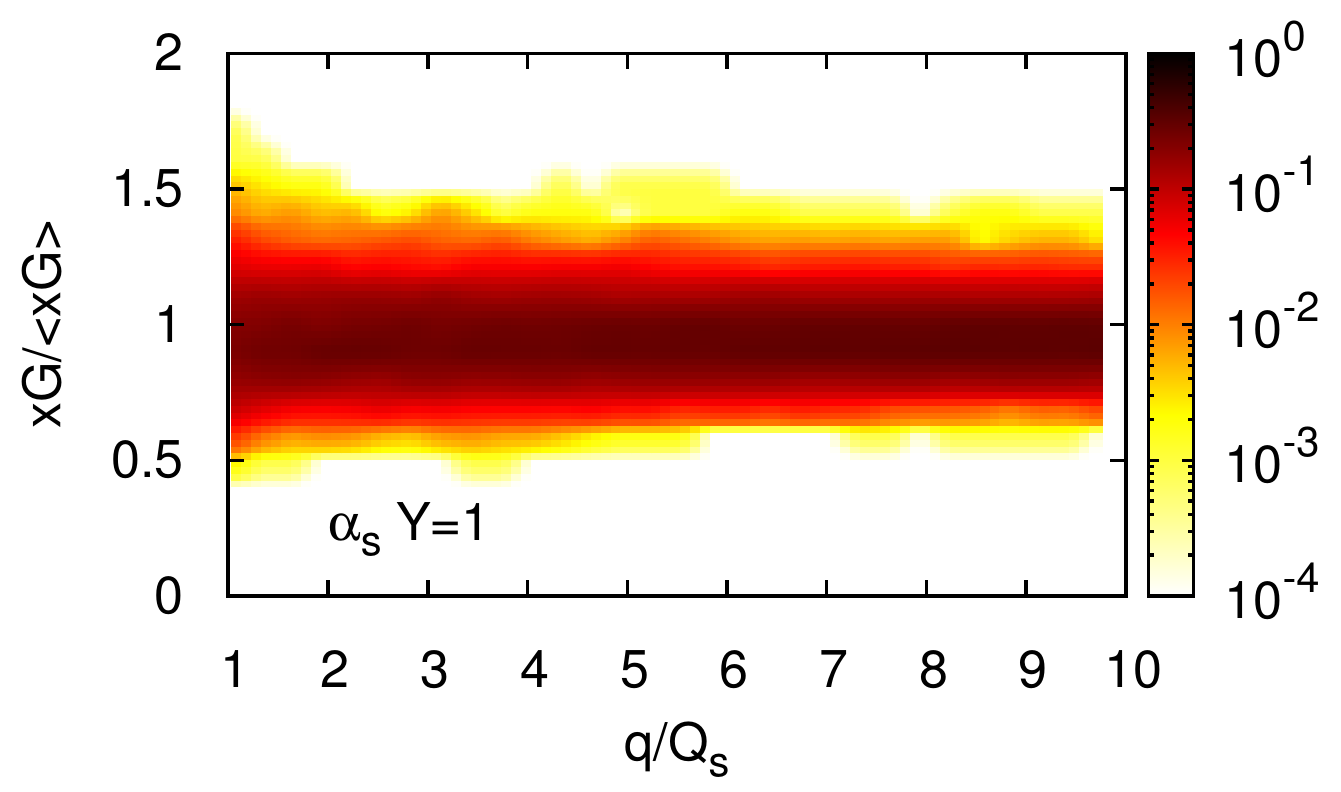}
\caption{Distribution of functions $xG^{(1)}(x,q^2)/\langle xG^{(1)}(x,q^2)\rangle$
  in the JIMWLK ensemble at $\alpha_s Y=1$. The color coding indicates
the probability for a particular function $xG^{(1)}(x,q^2)/\langle xG^{(1)}(x,q^2)\rangle$.}
\label{fig:xG-q_ensemble}       
\end{figure}

\begin{figure}[h]
\centering
\includegraphics[width=0.9\linewidth,clip]{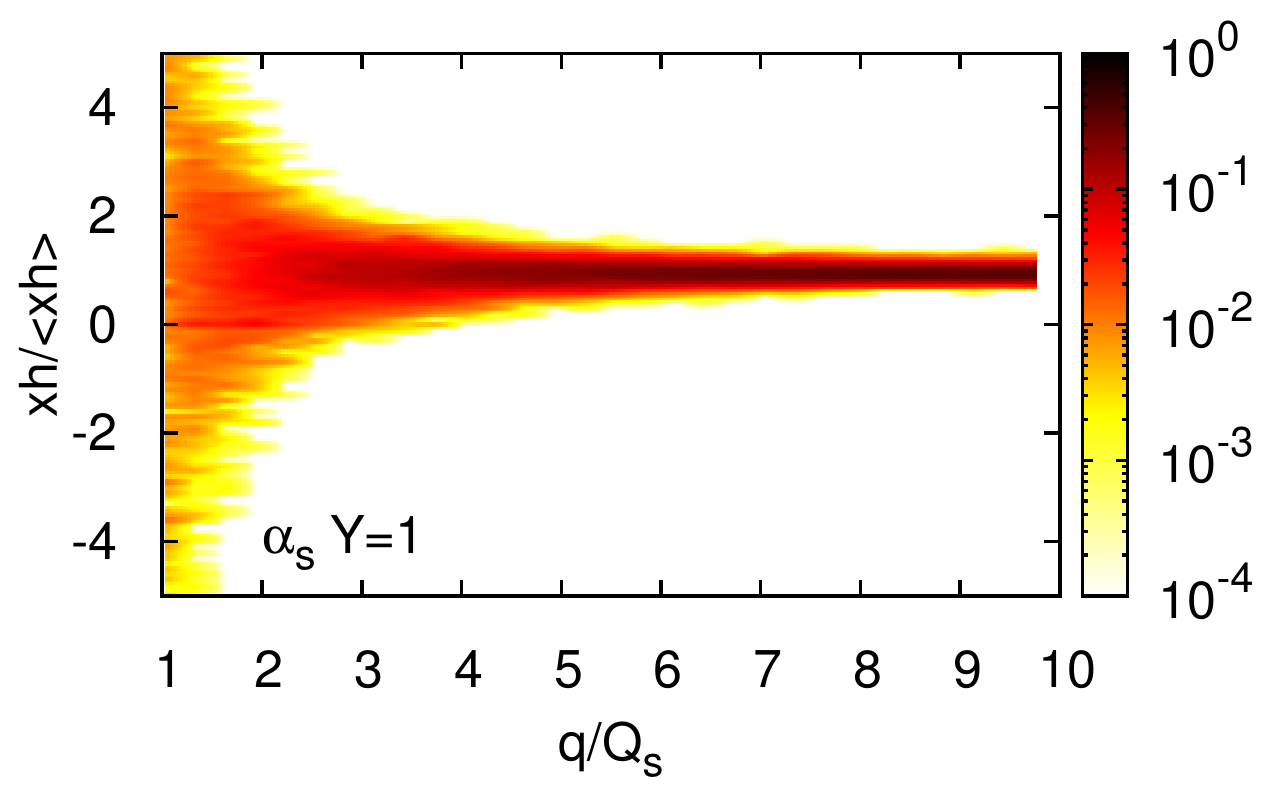}
\caption{Distribution of functions $xh^{(1)}(x,q^2)/\langle xh^{(1)}(x,q^2)\rangle$
  in the JIMWLK ensemble at $\alpha_s Y=1$.}
\label{fig:xh-q_ensemble}       
\end{figure}
The functional distributions of $xG^{(1)}(x,q^2)$ and
$xh^{(1)}(x,q^2)$ in the JIMWLK ensemble are shown in
figs.~\ref{fig:xG-q_ensemble} and~\ref{fig:xh-q_ensemble},
respectively. At high transverse momentum the distributions are
strongly peaked about the most likely WW functions. On the other hand,
when $q$ is not very far above $Q_s(Y)$ the ensemble of linearly
polarized WW gluon distribution functions is broad. At $\alpha_s
Y\sim1$ it includes non-positive definite functions as well
as functions which take values several times their average.

\section*{Acknowledgements}
A.D.\ thanks the organizers for the invitation to ISMD 2017; and
gratefully acknowledges support by the DOE Office of Nuclear Physics
through Grant No.\ DE-FG02-09ER41620, and from The City University of
New York through the PSC-CUNY Research grant 60262-0048.


\begin{thebibliography}{}
%
%

\bibitem{Aschenauer:2017jsk} 
E.~C.~Aschenauer {\it et al.},
arXiv:1708.01527 [nucl-ex].

\bibitem{Dominguez:2011wm} 
F.~Dominguez, C.~Marquet, B.~W.~Xiao and F.~Yuan,
Phys.\ Rev.\ D {\bf 83}, 105005 (2011)

\bibitem{Mulders:2000sh} 
P.~J.~Mulders and J.~Rodrigues,
Phys.\ Rev.\ D {\bf 63}, 094021 (2001)

\bibitem{Meissner:2007rx} 
S.~Meissner, A.~Metz and K.~Goeke,
Phys.\ Rev.\ D {\bf 76}, 034002 (2007)

\bibitem{Boer:2009nc} 
D.~Boer, P.~J.~Mulders and C.~Pisano,
Phys.\ Rev.\ D {\bf 80}, 094017 (2009)

\bibitem{Metz:2011wb} 
A.~Metz and J.~Zhou,
Phys.\ Rev.\ D {\bf 84}, 051503 (2011)

\bibitem{DQXY}
F.~Dominguez, J.~W.~Qiu, B.~W.~Xiao and F.~Yuan,
Phys.\ Rev.\ D {\bf 85}, 045003 (2012)

\bibitem{Boer:2010zf} 
D.~Boer, S.~J.~Brodsky, P.~J.~Mulders and C.~Pisano,
Phys.\ Rev.\ Lett.\  {\bf 106}, 132001 (2011)

\bibitem{Qiu:2011ai} 
J.~W.~Qiu, M.~Schlegel and W.~Vogelsang,
Phys.\ Rev.\ Lett.\  {\bf 107}, 062001 (2011)

\bibitem{Lansberg:2017tlc} 
J.~P.~Lansberg, C.~Pisano and M.~Schlegel,
Nucl.\ Phys.\ B {\bf 920}, 192 (2017);\\
J.~P.~Lansberg, C.~Pisano, F.~Scarpa and M.~Schlegel,
arXiv:1710.01684 [hep-ph].
  
\bibitem{Lappi:2017skr} 
T.~Lappi and S.~Schlichting,
arXiv:1708.08625 [hep-ph].

\bibitem{Kharzeev:2003wz} 
D.~Kharzeev, Y.~V.~Kovchegov and K.~Tuchin,
Phys.\ Rev.\ D {\bf 68}, 094013 (2003)

\bibitem{Dominguez:2011br} 
F.~Dominguez, J.~W.~Qiu, B.~W.~Xiao and F.~Yuan,
Phys.\ Rev.\ D {\bf 85}, 045003 (2012)
  
\bibitem{jimwlk}
J.~Jalilian-Marian, A.~Kovner, A.~Leonidov and H.~Weigert,
Nucl.\ Phys.\ B {\bf 504}, 415 (1997);
Phys.\ Rev.\ D {\bf 59}, 014014 (1998);\\
E.~Iancu, A.~Leonidov and L.~D.~McLerran,
Phys.\ Lett.\ B {\bf 510}, 133 (2001);
Nucl.\ Phys.\ A {\bf 692}, 583 (2001);\\
H.~Weigert,
Nucl.\ Phys.\ A {\bf 703}, 823 (2002)
  
\bibitem{Dumitru:2015gaa} 
A.~Dumitru, T.~Lappi and V.~Skokov,
Phys.\ Rev.\ Lett.\  {\bf 115}, no. 25, 252301 (2015)

\bibitem{Stasto:2000er} 
A.~M.~Stasto, K.~J.~Golec-Biernat and J.~Kwiecinski,
Phys.\ Rev.\ Lett.\  {\bf 86}, 596 (2001)
  
\bibitem{Iancu:2002aq} 
E.~Iancu, K.~Itakura and L.~McLerran,
Nucl.\ Phys.\ A {\bf 724}, 181 (2003);\\
J.~Jalilian-Marian and Y.~V.~Kovchegov,
Phys.\ Rev.\ D {\bf 70}, 114017 (2004);\\
H.~Fujii, F.~Gelis and R.~Venugopalan,
Nucl.\ Phys.\ A {\bf 780}, 146 (2006);\\
C.~Marquet and H.~Weigert,
Nucl.\ Phys.\ A {\bf 843}, 68 (2010);\\
E.~Iancu and D.~N.~Triantafyllopoulos,
JHEP {\bf 1204}, 025 (2012)

\bibitem{Dumitru:2016jku} 
A.~Dumitru and V.~Skokov,
Phys.\ Rev.\ D {\bf 94}, no. 1, 014030 (2016)

\bibitem{Marquet:2016cgx} 
C.~Marquet, E.~Petreska and C.~Roiesnel,
JHEP {\bf 1610}, 065 (2016)

\bibitem{Mueller:2002zm} 
A.~H.~Mueller and D.~N.~Triantafyllopoulos,
Nucl.\ Phys.\ B {\bf 640}, 331 (2002)
  
\bibitem{Dumitru:2017cwt} 
A.~Dumitru and V.~Skokov,
Phys.\ Rev.\ D {\bf 96}, 056029 (2017)

\bibitem{Kovchegov:1999ua} 
  Y.~V.~Kovchegov,
  Phys.\ Rev.\ D {\bf 61}, 074018 (2000)


\end{thebibliography}
\end{document}